\documentclass[pra,aps,12pt]{revtex4-1}
\usepackage{amssymb}
\usepackage{latexsym}
\usepackage{graphicx}
\usepackage{enumerate}
\usepackage{amsmath}
\usepackage{dcolumn}
\usepackage{bm}
\begin{document}

\author{Graciana Puentes}

\title{Unraveling the physics of topological phases \\ with random walks of light}

\begin{abstract}
I propose to study the complex physics of topological phases by an all optical implementation of a discrete-time quantum walk. The main novel ingredient is the use of parametric amplifiers in the random network which can in turn be used to emulate intra-atomic interactions and thus analyze many-body effects in topological phases even when using light as the quantum walker. I plan to characterize the intensity probability distribution and the spatial correlations of the output interference pattern for different input states, as well the robustness  of localized boundary states characterizing  topological insulators to different sources of noise. In particular, I expect to determine whether a non-local order parameter associated with a given topological entanglement measure can be determined in order to characterize topological order, and possible applications in entanglement topological protection. One of the most promising applications of the proposed research  includes the study of topological phases in photosynthetic energy transferring processes characterizing biological systems.%

\begin{flushleft}
{\small \textit{Keywords:} \textsf{Topological phases, quantum walks, quantum optics, quantum transport, biological processes}.}
\end{flushleft}

\end{abstract}

\maketitle
\section{INTRODUCTION}
\subsection{\emph{Quantum phase transitions}}


\noindent Phase transitions play a fundamental  role in physics. From  melting ice  to the creation of mass in the Universe, phase transitions are at the center of most dynamical processes which involve an abrupt change in the properties of a system.~Phase transitions are usually driven by some form of fluctuation. While classical phase transitions are typically driven by thermal noise, quantum phase transitions are triggered by  quantum fluctuations. Quantum phase transitions have been extensively studied in a large number of fields ranging from cosmology  to condensed matter and have received much attention in the field of ultra-cold atoms since the observation of Bose-Einstein condensation \cite{BEC}, and the subsequent experimental realization of Superfluid-Mott Insulator phase transition in optical lattices \cite{Greiner}. A common feature of quantum phase transitions is that they involve some form of spontaneous symmetry breaking, such that the ground state of the system  has less overall symmetry than the Hamiltonian and  can be described by a \emph{local} order parameter. \\

\noindent A rather distinctive class of quantum phases is present in systems characterized by a Hilbert space which is split into different topological sectors, the so called topological phases.  Topological phases have received much attention after the discovery of the quantum Hall effect \cite{Thouless} and the interest increased following the prediction \cite{Kane} and experimental realization \cite{Koening} of a new class of material called topological insulators. Topological insulators are band insulators with particular symmetry properties arising from spin-orbit interactions which are predicted to exhibit surface edge states which should reflect the non-trivial topological properties of the band structure, and which should be topologically protected by time reversal symmetry. Unlike most familiar phases of matter which break different kinds of symmetries, topological phases are not characterized by a broken symmetry, they have degenerate ground states  which present more symmetry than the underlying Hamiltonian, and can not be described by a \emph{local} order parameter. Rather, these partially unexplored type of phases are described by topological invariants, such as the Chern number which is intimately related to the adiabatic Berry phase, and are predicted to convey a variety of exotic phenomena, such as fractional charges and magnetic monopoles \cite{Qi}. It has recently been theoretically demonstrated that it is possible to simulate a large ``Zoo" of topological phases by means of discrete time quantum walks (DTQW) of a single particle hopping between adjacent sites of an optical lattice, through a sequence of unitary operations \cite{Kitagawa}.

\subsection{\emph{Quantum walks}}

\noindent Random walks have been used to model a variety of dynamical physical processes containing some form of stochasticity, including  phenomena such as Brownian motion and the transition from binomial to Gaussian distribution in the limit of large statistics. The quantum walk (QW) is the quantum analogue of the random walk, where the classical walker is replaced by a quantum particle, such as a photon or an electron, and the stochastic evolution is replaced by a unitary process. The stochastic ingredient is added by introducing some internal degrees of freedom  which can be stochastically flipped during the evolution, which is usually referred to as a  quantum coin. One of the main ingredients of quantum walks is that the different paths of the quantum walker can interfere, and therefore present a complicated (non Gaussian) probability distribution for the final position of the walker after a number of  discrete steps. In recent years, quantum walks have have been successfully implemented to simulate a number of quantum phenomena such as photosyntesis \cite{Sension,  Mohseni}, quantum diffusion \cite{Godoy}, vortex transport \cite{Rudner} and electrical brake-down \cite{Oka}, and they have provided a robust platform for the simulation of quantum algorithms and maps \cite{Paz}. QW have been experimentally implemented in the context of NMR \cite{Ryan}, cavity QED \cite{Agarwal}, trapped ions \cite{Ions}, trapped neutral atoms \cite{Karski} as well as  optics, both in the spatial \cite{Do} and frequency domain \cite{Bouwmeester}. In recent years, quantum walks with single and correlated photons have been successfully introduced using wave-guides \cite{Peruzzo} and bulk optics \cite{White}. It is relevant to point out that any implementation of a quantum walk so far \cite{Godoy, Rudner, Oka, Paz, Agarwal, Karski, Bouwmeester, Peruzzo, White} has introduced passive linear elements \emph{only} for the composing elements of the random network. 

\subsection{\emph{Topological Insulators}}

\noindent Topological insulators is a novel kind of solid which are predicted to exhibit surface states leading to quantized conductance of the carriers. The surface state of topological insulators is closely related to the Dirac electronic structure of Graphene, which has a linear energy-momentum relationship (known as Dirac cone)  \cite{Moore}. The corresponding surface (2D) or edge (1D) states in topological insulators have helical properties which interconnect spin and momentum. Unlike quantum Hall effect where a magnetic field induces cyclotron motion of electrons which is essential for the formation of edge states, the observation of helical edge states characterizing topological insulators does not require application of a magnetic field \cite{Hasan}. A full class of topological insulators can be realized in a system of non-interacting particles, with a binary (psuedo) spin space for (bosons) fermions, via a random walk of discrete time unitary steps as described in Ref \cite{Kitagawa}. The particular type of phase is determined by the size of the system (1D or 2D) and by the underlying symmetries characterizing the Hamiltonian, such as particle-hole symmetry (PHS), time-reversal symmetry (TRS), or chiral symmetry (CS). The 1D discrete time quantum walk (DTQW) can be specified by a series of unitary spin dependent translations $T$ and rotations $R(\theta)$, where $\theta$ specifies the rotation angle. Thus, the quantum evolution is determined by applying a series of unitary operations or steps:

\begin{equation}
\label{eq:1}
U(\theta)=TR(\theta).
\end{equation}

\noindent The spin dependent translation $T$ is given by:

\begin{equation}
T=\sum_{j}|j+1\rangle\langle j | \otimes|\uparrow\rangle \langle \uparrow| +|j-1\rangle \langle j| \otimes |\downarrow\rangle \langle \downarrow|,
\end{equation}

\noindent where the index $j$  determines wether the quantum walker hops to the left or right depending on the outcome of a binary stochastic coin. The spin rotation $R(\theta)$ around an axis perpendicular to the direction of the displacement  by an angle $\theta$ is given by:

\begin{equation}
R(\theta)=
\left( {\begin{array}{cc}
 \cos(\theta/2) & -\sin(\theta/2)  \\
 \sin(\theta/2) & \cos(\theta/2)  \\
 \end{array} } \right).
\end{equation}

\noindent The generator of the unitary evolution operator (or map) in Eq. [1] is the time-independent Hamiltonian $H(\theta)$, for which the discrete time evolution operator  $U(\theta)$ can be defined as:

\begin{equation}
U(\theta)=e ^{-iH(\theta)\delta t},
\end{equation}

\noindent where we have chosen $\hbar=1$, and the finite time evolution after $N$ steps is given by  $U^{N}=e^{-i H(\theta)N\delta t}$. \\

\noindent The Hamiltonian $H(\theta)$ determined by the translation and rotation steps $T$ and $R(\theta)$, posses particle hole symmetry (PHS)  for some operator $P$ (i.e. $PHP^{-1}=-H$) and it also contains chiral symmetry (CS). The presence of PHS and CS guaranties time reversal symmetry (TRS). The presence of TRS and PHS imply that the system belongs to a topological class contained in the Su-Schrieffer-Heeger (SSH) model \cite{Su} and can thus be employed to simulate a class of SSH topological phase.~An extension to 2D topological insulator can be obtained by extending the lattice of sites to 2D. Different geometries such as square lattice or triangular lattice are described in Ref \cite{Kitagawa}.

\subsection{\emph{Parametric amplification and} SU(1,1) \emph{symmetry}}

\noindent It should be noted that by changing the unitary step $T$ and $R(\theta)$ it is possible to simulate other types of topological phases. In particular, it is interesting to note that the matrix $R(\theta)$ defined in Eq (3)  belongs to the SU(2) symmetry class, which correspond to linear passive transformations. In optical terms, such devices may comprise half wave-pates, polarizers or beam splitter. However, the most general class of quadratic transformations should also include SU(1,1) matrices, which describe active operations. Among this category we find parametric amplifiers, four-wave mixers, and phase conjugating mirrors. Thus a more general evolution is given in terms of a parametric rotation matrix $R(\chi)$, given by:

\begin{equation}
R(\chi)=
\left( {\begin{array}{cc}
 \cosh(\chi) & i\sinh(\chi)  \\
 i\sinh(\chi) & \cosh(\chi)  \\
 \end{array} } \right),
\end{equation}

\noindent where $R(\chi)$ belongs to SU(2) class for $\chi$ imaginary and involves only passive operations. For $\chi$ real the parametric rotation belongs  to the SU(1,1) class and includes active transformations, as characterized by parametric amplifiers. We will denote such general evolution $U_{ge}(\chi)=TR(\chi)$, and as it will be described in the next section it will be the main ingredient for our proposed work. 


\vspace{0.2cm}

\section{PROPOSED RESEARCH WORK}

\subsection{\emph{All optical simulation of topological insulators}} 

\noindent The transformation described in the previous sections for the simulation of topological phases with non-interacting particles requires only a conditional translation step and a rotation step, in a system with a 2-level internal degrees of freedom which  plays the role of the quantum coin. Therefore, it is directly amenable to experimental realizations with single photons if we identify polarization as the  2-level quantum system of interest. Furthermore, due to the analogy between single photons and classical states the same scheme can also be implemented using classical benchmark states. While the translation step described in Eq. (2) can be successively implemented by means of a network of polarizing  beam splitters cubes \cite{White}, or fibers \cite{Peruzzo}; the polarization rotation can be accomplished by means of half wave-plates in the bulk case, or by controllable bending-induced birefringence in fibers \cite{GPuentes3}. In this work I propose to realize a 1D and 2D quantum walk for the simulation of topological phases using a network of fiber beam-splitters.~The main step in the quantum walk is characterized by the unitary evolution operator:

\begin{equation}
U_{ge}(\chi)=TR(\chi),
\end{equation}

\noindent where $R(\chi)$ is a matrix with SU symmetry, and the finite time evolution is given by $U_{ge}(\chi)^N=e^{-iH(\chi)N\delta t}$. Depending on the underlying symmetries (i.e. PHS, TRS or CS) in $H(\chi)$ it is possible to implement a full class of topological phases.\\

\noindent A distinctively different kind of dynamics can be observed depending on wether $\chi$ is real or imaginary. As it was pointed out in \cite{Torma}, for $\chi \in \Im$, the rotation step has SU(2) symmetry and corresponds to a passive transformation, which should only relate to diabatic and adiabatic behavior in the network, depending on the actual value of $\chi$. This dynamics can in turn be related to Landau Zener transitions at each network crossings. On the other hand, for  $\chi \in \Re$, the rotation belongs to SU(1,1) and  we expect to encounter an active  transformation in the system.  Such form of amplification is, in turn, related to quantum and classical transitions in the network. In particular,  the now 2-mode input state can introduce quantum correlations in the system, and it is interesting to note that in this case there should be sensitivity to phases in the input states, as well as some critical value of the gain or amplification, above which no quantum effect should be observed \cite{Torma}. This kind of non-linear term has  an analogue in the atomic world in terms of intra-particle interactions. Therefore, by tailoring the amplification it is possible to simulate different many-body Hamiltonians, and the quantum-classical transition due to decoherence effect in the many-body states. Further investigations of the amplifying network include effects of different 2-mode input states with different kinds of symmetries and correlations, such as phase and number squeezed states, or quadrature entangled states.\\

\subsection{\emph{Detection: Split sets and edge states}}

\noindent In Ref \cite{Kitagawa}, the authors propose to detect the existence of the non-trivial topological order by introducing a boundary between distinct phases, where a localized edge state should be found. To this end they introduce an additional ``split-step" in the quantum walk such that the translation for the spin up ($\uparrow$) and down ($\downarrow$) are separated by an additional rotation by an angle $\theta_{2}$. In this way, each step in the quantum walk can be written as:

\begin{equation}
U(\theta_{1},\theta_{2})=T_{\downarrow}R(\theta_{2})T_{\uparrow}R(\theta_{1}),
\end{equation}

\noindent where each translation $\{T_{\uparrow},T_{\downarrow}\}$ shifts the position to either left or right by one lattice site, and $R(\theta_{2})$ is a site dependent rotation by and angle $\theta_{2}(x)$. By changing the value of $\theta_{2}$ along the lattice position $x$ it is possible to realize different topological phases characterized by a different integer winding number (analogue to the quantum Berry phase) and thus cross the boundary between different topological sectors where at least one localized state should be found. While, on the other hand if $\theta_{2}$ remains constant, there is a single topological domain and we should expect no localized edge states. In the case of using a single mode input beam in a coherent state, we expect to find a peak centered at the boundary between the two topological regions following Poissonian count statistics, while in the case of no topological boundary the statistics should be binary along the 1D lattice direction, once the wave functions of the initial state has spread out sufficiently ($N>>1$). It is worth to point out that such localized states are topologically protected, so it would be of great interest to test the robustness of such states to different kinds of noise such as amplitude noise or frequency noise in the source.  \\

\noindent When using non-linear amplifying elements in the network it will be of interest to prepare 2-mode input states with different kinds of statistics and fluctuations, such as quadrature squeezed states. The characterization of the output signal should not only be performed in space, where a binary or Subpoissonian statistics of counts should be observed for a gain smaller or larger that some critical value $G_{c}$, but also in the correlations of the field. The field spatial correlations can in turn be determined by measuring the single-mode (two-mode) correlation function $g^{1(2)}$. It would be of great relevance to find out wether entanglement increases or decreases along the network, and wether 2-mode boundary states present any kind of squeezing and are robust against different kinds of perturbations, such amplitude noise or phase noise. It would also be interesting to find out if there are any decoherence free input states, which are topologically protected against decoherence \cite{PuentesJPB2012}. Finally, it remains of both theoretical and experimental interest to find out if a \emph{non-local} order parameter associated with some entanglement measure can be defined in order to characterize the topological phase.

\subsection{\emph{ Energy transferring processes in biological systems}}

\noindent Energy transferring processes involving a single excitation can be modeled by means of a Master equation describing a quantum trajectory in  an open system \cite{Sension, Mohseni}. The quantum trajectory is in turn subject to random quantum jumps and to decoherence due to coupling with a phonon bath. In optical terms, it is possible to implement such quantum dynamics by means of a quantum walk  in some form of active medium, which should be resonant with the frequency of the input light. In this way the quantum jumps can be modeled by a series of absorptions and spontaneous emissions of photons by the atomic medium, though we anticipate that the modeling of such quantum jumps  would remain part of a future efforts. The phonon bath can in turn be simulated by the transverse spatial modes of the radiation field in such way that by averaging over the spatial modes it is possible to simulate decoherence due to coupling to an external bath \cite{GPuentes2}. Furthermore, contrary to intuition, it has recently become apparent that quantum transport in biomolecules can be enhanced by dephasing mechanisms induced by a noisy enviroment \cite{Plenio}. It would be of great interest to experimentally investigate the role of noise in transport phenomena and in topologically protected subspaces simulated by means a quantum walk.

\subsection{\emph{Aims of the proposed research}}

\noindent The main purpose of the proposed research can be outlined in the following tasks:

\begin{itemize}

\item {\bf Simulation and detection of topological phases via all optical 1D and 2D random walks} \\
\noindent Can we experimentally simulate 1D and 2D topological phase transitions with coherent states of light via all optical random walks in a fiber network? Is there a way to characterize such phases not only by the introduction of edge states, but also by measuring the statistics of the field by means of intensity measurements and spatial field correlations?

\item{\bf Simulation of many-body effects in topological phases by non-linear parametric amplification in fiber networks}\\
Is there any non-trivial effect by introducing amplification (i.e. non-linearity) in quantum walks? Can this non-linearity simulate intra-particle interactions and be used for simulation of many-body physics in bosonic systems? Is there a different behavior of the amplifying network for 2-mode input states with different kind of statistics and fluctuations? What is the role of decoherence in amplifying networks and many-body topological phases?

\item{\bf Application: Topological  order  in energy transferring biological process}\\
A potential application of this work is in the study of topological order in biological processes. In particular it would be of great interest to analyze the existence of topologically protected states in photosynthetic energy transferring processes, such as light-harvesting complexes of higher plants \cite{Novo, Sension}, which can in turn be simulated by means of a continuous time random walk  with non-unitary  dynamics \cite{Sension, Mohseni}.  Furthermore, I also plan to analyze the role of  amplification and correlations in topologically protected states and energy transferring efficiency mechanisms characterizing such dynamical biological processes. 


\end{itemize}


\section{EXPERIMENTAL METHODS}

\subsection{\emph{Fiber Network}}

\noindent In Ref \cite{White} the authors performed an optical implementation of the operator defined by Eq. (2), using polarization degrees of freedom of single photons and a sequence of half-wave plates and calcite beam-splitters. On the other hand, in Ref \cite{Peruzzo} the authors implemented a quantum walk in a lattice of coupled wave-guides. 
In this work I propose to use a fiber network to implement a quantum walk to simulate 1D and 2D  topological insulators, using coherent states and squeezed coherent states of light with Poissonian and Subpoissonian  statistics, respectively. Non-classical effects are expected in the latter case. \\

\noindent One of the main ingredients is the implementation of an optical non-linearity which can introduce amplification, and simulate in this way both attractive and repulsive interactions. A similar idea was already proposed in \cite{Silberberg}, where attractive interactions were introduced in a planar AlGaAs waveguide characterized by  a strong focussing Kerr non-linearity. Likewise, repulsive interactions can be simulated using defocusing non-linearities \cite{Silberberg}, though this would remain part of future efforts. A suitable alternative kind of waveguide for the simulation of attractive interactions are photonic band gap fibers with a Raman active gas, which are predicted to have a strong non-linearity. These fibers consist of a hollow core photonic crystal fiber filled with an active Raman gas which are capable of exceeding intrinsic Kerr non-linearities by orders of magnitude   \cite{Skryabin}.  

\subsection{\emph{Input state preparation}}

\noindent For the non-interacting case (SU(2) symmetry) I plan to use single-mode states both with Poissonian and Subpoissonian  statistics, such as coherent states and squeezed coherent  states. The non-classical nature of the squeezed states should be revealed in the intensity distribution of counts as well as in the standard deviation. In particular, I expect to find localized edge states in the case of light with non-classical statistics only. Additionally, we plan to introduce controlled amplitude or frequency noise in order to test the robustness of the edge state.\\

\noindent For the interacting case (SU(1,1)  symmetry) I plan to use also 2-mode input states with quadrature correlations, such inputs states can be produced in non-linear crystals by the process of spontaneous parametric down conversion \cite{Mosley}. Quantum theory then predicts that the probability amplitudes of the modes should interfere leading to an enhancement/reduction of the initial correlations. One of the goals of the project is to analyze the sensitivity of the non-linear network to phase relations in the input state and to the amount of gain. I also plan to analyze the effect of correlations and entanglement in the input state on localized edge states and to find some kind of non-local order parameter characterizing topological order \cite{Haldane}. 

\subsection{\emph{Detection Schemes}}

\begin{itemize}
\item{\bf Intensity probability distributions and standard deviation}\\
\noindent The most direct form of measurement is to detect the statistics of counts by studying intensity histograms and their standard deviation, as described in Ref \cite{White, Silberberg}. In particular, by placing a photo-diode at the output of each fiber,  characterizing a given site in the network, it is possible to obtain a probability distribution of counts and its standard deviation along the $N$ steps of the quantum walk.  While in the case of input states with Poissonian statistics I expect to find a classical \emph{binary} distribution of counts as the output of the quantum walk, in the non-classical case we expect to find a localized edge state at the boundary between two different topological sectors. Furthermore, I plan to measure the normalized standard deviation $\sigma_{N}$ for the classical and non-classical case, where we expect to find a markedly different dependance on the number of steps $N$; namely, while for the classical walk (coherent states)  I plan to obtain   $\sigma_{N} \propto \sqrt{N}$ dependance, for the non-classical case  (squeezed states) I plan to obtain an $\sigma_{N} \propto N $ dependance with the number of discrete steps.

\item{\bf HBT correlation measurements}\\
\noindent When using non-linear fibers in the amplifying network, it would be interesting to analyze 1-mode ($g^{1}(\Delta r)$) and 2-mode ($g^{2}(\Delta r)$) spatial correlations functions by means of Hanbury-Brown-Twiss (HBT) like interferometers between different output modes in the network, as described in Ref \cite{Silberberg}.  In particular, while in the case of attractive interactions, as simulated by Kerr non-linearities in fibers, the correlations are expected to increase, for the repulsive case the correlations are expected to decrease. I also plan to analyze the dependance of spatial correlations  on the amount of gain present in the medium. In particular, for some critical value of the overall gain $G_{c}$ I expect to find a decay of the correlations, which in turn can be related to the classical-quantum transition in amplifying media \cite{Torma}. Finally, I will also investigate the response of the amplifying network to different phases in the input states, as well as to phase noise, by introducing phase averaging mechanisms.

 \end{itemize}
 
 \vspace{1cm}
 
\section{TIMELINE}

\noindent Bellow there is a table summarizing the main goals of the project and an estimate of the time required to accomplish the research work pertaining to each goal.  Please note that these timelines are only indicative and can be accommodated during the project if required. Also, other additional research lines might be included during the course of the work. 

\newpage
\begin{table}[h!]
\begin{center}
\begin{tabular}{|c||l||l|}
\hline
\hline
Months & \hspace{2.4cm} Subject & \hspace{3cm} Details \\
\hline
\hline
$1$ to $6$ & 1D and 2D QW with SU(2) symmetry & Optical implementation via fiber network\\
&Topological invariants and edge states &  Detection of boundary states by intensity histograms\\
\hline
$6$ to $12$ & 1D and 2D QW with SU(1,1) symmetry & Measurement of correlation functions    \\ 
&Many-body dynamics & Dependance on gain and phase noise \\
\hline
$12$ to $18$ & Topological phases in QW and& Continuous time QW and non-unitary operations\\ & study of energy transferring processes & Topological order and (spatial) decoherence in biology\\
\hline
\hline
\end{tabular}
\end{center}
\end{table}

\section{SUMMARY AND OUTLOOK}

\noindent In this work I propose an experimental implementation of topological phases by means of optical fiber networks. I expect to characterize the robustness of such topological phases and their characteristic boundary states against amplitude and phase noise as well as to decoherence, by tracing over spatial modes of the field. One of the main goals of the proposed work is to investigate the use of parametric amplifiers as a means of simulating many-body effects in topological phases. In particular, I expect to link such phases with the classical or quantum statistics of the fields by means of intensity distribution and spatial correlation measurements, and I intend to find a link between some measure of entanglement and a non-local order parameter characterizing the topology of the phase \cite{Haldane}, or the feasibility entanglement topological protection \cite{PuentesJPB2012} Some significant signatures of many-body dynamics in topological order are expected to be apparent, such as charge fractionalization and Hall quantization, which motivate the extension of the research to the non-linear (many-body) scenario. Furthermore, other more complex topological phases (such as spin Hall phase) could  be simulated in the future by all optical means by using 2D random walks and higher dimensional internal degrees of freedom of the radiation field, such as  the orbital angular momentum \cite{PuentesPRL2012}. One of the main future direction of this work is in biophysics and the study of energy transfer efficiency in  dynamical processes, where some of the most fundamental processes such as photosynthesis are currently explained in terms of quantum walks with non-unitary operators  \cite{Mohseni}. It would be of great relevance to find out if biological systems indeed provide any form of topologically protected states or mechanisms.  Furthermore, topological order has been  considered as a useful ingredient  for fault tolerant quantum computation, as it can protect the system against local perturbations which would otherwise introduce decoherence and loss of quantum information  \cite{Preskill}. 

\section*{AKNOWLEDGEMENTS}

The main ideas and concepts included in this Research Program were ellaborated by the author during the period 2010-2011.

\end{document}